\documentclass[a4paper,fleqn]{cas-dc}
\usepackage[numbers]{natbib}
\usepackage{caption}
\usepackage{amsmath}
\usepackage[scr]{rsfso}
\newcommand{\Laplace}{\mathscr{L}}
\usepackage{algorithm} 
\usepackage{algorithmic}  
\usepackage[algo2e]{algorithm2e} 
\SetKwInput{KwInput}{Input}                
\SetKwInput{KwOutput}{Output}

\renewcommand{\thesection}{\arabic{section}}

\begin{document}
\let\WriteBookmarks\relax
\def\floatpagepagefraction{1}
\def\textpagefraction{.001}

\shorttitle{Decentralized Accessibility of \textit{e}-commerce Products}

\shortauthors{Gulshan Kumar, Rahul Saha, William J Buchanan, G. Geetha, Reji Thomas, Mritunjay Kumar Rai, Tai-Hoon Kim}

\title[mode = title]{Decentralized Accessibility of \textit{e}-commerce Products through Blockchain Technology}

\author[1]{Gulshan Kumar}
\cormark[1]
\ead{gulshan3971@gmail.com}
\address[1]{School of Computer Science and Engineering, Lovely Professional University, India}

\author[1]{Rahul Saha}

\author[2]{William J Buchanan}
\address[2]{Blockpass ID Lab, Edinburgh Napier University, UK}

\author[1]{G. Geetha}

\author[1]{Reji Thomas}

\author[3]{Tai-Hoon Kim}
\address[3]{School of Economics and Management, Beijing Jiaotong University, China}

\author[4]{Mamoun Alazab}
\address[4]{Charles Darwin University, Australia}

\cortext[cor1]{Principal corresponding author}

\markboth{Gulshan Kumar, et al}{PRODCHAIN}

\begin{abstract}
A distributed and transparent ledger system is considered for various \textit{e}-commerce products including health medicines, electronics, security appliances, food products and many more to ensure technological and e-commerce sustainability. This solution, named as 'PRODCHAIN', is a generic blockchain framework with lattice-based cryptographic processes for reducing the complexity for tracing the e-commerce products. Moreover, we have introduced a rating based consensus process called Proof of Accomplishment (PoA). The solution has been analyzed and experimental studies are performed on Ethereum network. The results are discussed in terms of latency and throughput which prove the efficiency of PRODCHAIN in \textit{e}-commerce products and services.The presented solution is beneficial for improving the traceability of the products ensuring the social and financial sustainability. This work will help the researchers to gain knowledge about the blockchain implications for supply chain possibilities in future developments for society.             
\end{abstract}

\maketitle

\section{Introduction}
Smart technologies and fast communication have significantly enhanced the product and service sectors in recent years. Organizations and enterprises are developing new things for product improvements. Multiple organizations are at present realizing assembled products through collaboration, for example car and computer manufacturing. Moreover, the cost and usability create 'demand and supply' a recurring process and thus 'product and supply' chains are closely connected  \cite{wong2019critical}\cite{REDDYK201940}. Each phase in the development of a product or service is considered in the product chain, whereas supply chain deals with the transaction of the developed product from the manufacturer to the customer. It means that if the product chain is maintained and monitored well, management of supply chain will become easier for the concerned organizations. The intrinsic risk associated with the data of such chains is worth considering. Generally, financial institutions apply traditional credit rating models to assess the credit risk of a company and often fail to provide an absolute assessment for small and medium companies/enterprises. On the other side, buyers assess the suppliers by valuing comprehensive vendor ratings on a broad range of operational performance. However, there are ways to integrate the financial and vendor ratings to a supply chain credit rating model that considers the financial enablers of the supplier and its operational evaluation provided by buyers jointly \cite{moretto2019supply}. The supply chain also faces risks from sustain-ability-related factors. Economic, social and environmental are  the base for evaluating the sustainability risk of a supply chain.   An aggregated metric is much more beneficial in this aspect of risk and material quality assessments  \cite{xu2019supply}. The bullwhip effect introduces uncertainty in the chain rule and complete development of a product passes through several risk assessment processes \cite{braz2018bullwhip}. However, there is no such online process available to monitor these phases of development in a decentralized fashion. Moreover, with the advent of IoT, big data and cloud computing and industry 4.0, the generated data in supply chain management or product management lines is a knowledge resource for improving the products of the future markets \cite{schniederjans2019supply}. 

The traditional supply chain management system is confined to a state or a country in earlier times. But, with the international diplomatic relationships and import-export provisions, product chain and supply chain- both have been extended to inter and intra continents. This concept of global supply chains has raised the bar of economic valuations in every country and therefore gathers great attention \cite{koberg2018systematic}. Along with the economic benefits, product chain and supply chain can be mismanaged if illegal and unauthorized access have been not prevented. According to the report of Kroll, 42\% of all the global supply chain and product chain management companies have faced at least one fraud incident in the last 10-15 years. Though Information Technology (IT) can perform the detection and prevention of fraud and other supply chain nightmares, the report suggests "it can also be a huge threat to a company's operations, reputation and future business prospects" \cite{Wailgum}. These frauds are broadly categorized as: financial, misinterpretation of goods or services, bribery including Foreign Corrupt Practices Act (FCPA) violations, sanctioning violation and kickbacks \cite{Pomerantz}.

The supply chain has been rigorously applied in various applications ranging from agriculture-food products \cite{behzadi2018agribusiness}\cite{jonkman2019integrating} to the vaccination \cite{duijzer2018literature}, fashion apparels \cite{wen2018fashion}, wind turbines \cite{megahed2018tactical} and even in healthcare \cite{Khos2019}. All of these applications are having a strong correlation with intermediaries as an integral part of supply chain. However, they do not possess a direct impact on product chain. With the untrusted intermediaries and malicious perspective, the significance of supply chain and product chain is viable to be demolished easily.  To sustain the chain rule without intermediaries and bridging the gap between producer and consumer,  an approach is needed in the present global business environment \cite{cole2019role}. The much needed transparent approach should be beneficial for social and economic sustainable developments. Therefore, a blockchain-based solution that integrates the product chain and supply chain to provide a transparent and decentralized resource of product and its access information has been proposed in the present communication.   This global platform, named as "\textit{PRODCHAIN}", is implicitly beneficial for e-commerce world. E-commerce has been a major enabler in financial growth and social perspectives in recent years. PRODCHAIN objectifies to enhance the attributes of e-commerce through improving transactions in supply chain management. The major contributions in the paper are:

\begin{itemize}
\item     Merging of value chain and supply chain into a single transparent blockchain-based solution,
\item     The use of blockchain for e-commerce with the start from product development to customer acquisition
\item     Lattice-based cryptography usage in blockchain signcryption process. 
\item     Introduction of Proof-of-Accomplishment as an extension of smart contract in heterogeneous stakeholder environment. 
\end{itemize}

The rest of the paper has been segregated in four parts. Section 2 deals with some basic concepts and blockchain applicability of product chain and supply chain. Section 3 conceptualizes the proposed solution which has been analyzed with experimentally and theoretically in section 4. Finally, section 5 concludes the work and shows future direction.

\section{Related Work}
\begin{figure*}
\centering
  \includegraphics[width=0.8\linewidth]{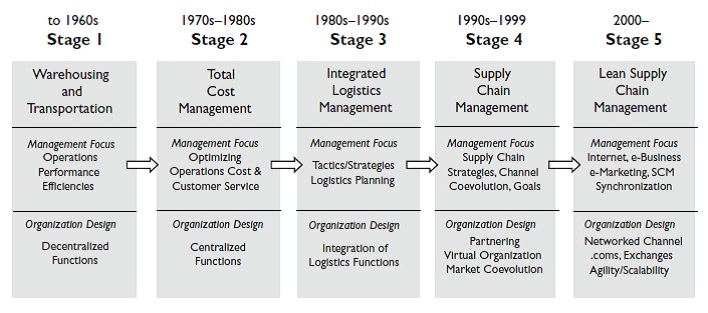}
  \caption{Evolution of Supply Chain Management \cite{daud2011lean}}
  \label{fig:fig01}
  \centering
\end{figure*}

The production chain is defined as an analytical tool to assess the production processes of both goods and services and their transformations \cite{Drahokoupil}. Generally, production process deals with sequence of activities with an output to an end product i.e. a chain of linked functions. Each stage in this process adds value to the production sequence and that is why "product chain" is often called "value-added" or "value" chains, interchangeably used in this communication. In this present century, technological shift towards IoT, cloud, blockchain and the liberalization of trade have radically reorganized the production processes as sliced-up pieces. This ensures in-depth analysis in each segment possible and efficient. While product chain is specific of a good or service, supply chain deals with the operational perspective including product development (product chain), marketing, operations, distributions, finance and customer service \cite{Tarver}. The evolution of the supply chain has been shown in Figure \ref{fig:fig01}.

A report by World Trade Organization (WTO) and co-published by the World Bank Group, the Organisation for Economic Co-operation and Development (OECD), the Institute of Developing Economies (IDE-JETRO) and the Research Center of Global Value Chains of the University of International Business and Economics (UIBE) confirms that global value chains have impacted positively on more than two-thirds of world trade in recent years  \cite{World}. It has been also urged in the report that governments require to promote the provisions that are conducive to investment, build the overall skills of local manufacturers and emphasize relationships among technology providers and local producers. Therefore, the need for a transparent and decentralized value chaining process and supply chain process has always been felt by the organizations. Thus, blockchain technology finds its way into value chain and supply chain to prevent all authentications and malicious frauds.  

The work executed in initializing the feasibility of blockchain in a supply chain management system is only a few. Recent work in this dimension affirms that a blockchain-based knowledge framework enhances assurance and reduces the perceived risks \cite{montecchi2019s}\cite{Alam2020}. Smart contract-based product traceability has been researched to re-engineer the product which is directly connected to the product chain too \cite{chang2019supply}. The use of permissioned and open blockchain in distribution industry has been reported \cite{banerjee2018blockchain}. Risk analysis in global supply chain by measuring the mean-variance is also an addition in this direction of work \cite{choi2019mean}. Food traceability \cite{choi2019blockchain}, luxury supply \cite{behnke2019boundary} and agriculture supply chain \cite{kamble2019modeling} are identified as the benefits of the field with blockchain implications in maintaining the business processes. Furthermore, the multidimensional relationship between blockchain and its feasibility in supply chain has been rigorously researched in recent times \cite{AZZI2019582}\cite{wang2019making}.  Along with supply chain management, blockchain technology has also been included in the family of Industry 4.0 to confirm data integrity and avoid tampering. Moreover, it is also beneficial against the problem of centralized single point failure by providing fault-tolerance, immutability, trust, transparency and full traceability of the stored transaction records. Blockchain research has also ensured such benefits in agri-food value chain partners in recent time \cite{zhao2019blockchain}. Industrial Product-Service Systems (IPS2) have also identified the scope of blockchain inclusion for monitoring the chained processes in a product life-cycle \cite{huang2019use}. Recently, work on blockchain-based logistics application for easy service delivery and assured decentralized resource for \textit{e}-commerce has been significantly researched \cite{li2019blockchain}. The authors have used Proof-of-Concept (PoC) and claims for providing non-repudiation, fairness, and confidentiality. 

The exploration of blockchain adaptability in supply chain management has been observed in \cite{Lu2017}, \cite{Jabbari} and \cite{Dujak2019}. Some more recent works for feasibility of blockchain in supply chain have explored in \cite{Ven2020}\cite{Longo2019}\cite{Helo2019}. The greening of blockchain involvement in supply chain is well discussed with pros and cons in \cite{Kou2018}. The implementation of blockchain in supply chain management is defined by Asia and specifically by China as shown in \cite{Kshetri2018},\cite{Ying2018} and \cite{Tian2016}.  Furthermore, the blockchain attributes and transparency for supply chain management are discussed on some recent works \cite{Niko2018} and \cite{KimShin2019}.  The design of an ontology-based provenance mechanism for blockchain in supply chains has been observed in literature \cite{Las2018}. A case study-based discussion for ontology driven blockchain properties shows the futuristic development feasibility of blockchain designs. Blockchain based supply chain applications have been studied in \cite{Wang2020}\cite{Helo2020}\cite{Liu2020} and \cite{Chen2020}.

The above analysis summarizes that supply chain management is in rigorous process of research for exploring the blockchain potentials.  The application of blockchain in supply chain is prominent; however, its feasibility in the product chain including value and supply is a step back and has not been explored significantly. Therefore, the proposed approach provides a solution based on blockchain that integrates both the value chain and supply chain with blockchain lattice. The features blockchain concepts used in the proposed model of operation is given as follows \cite{leewayhertz}:

\begin{itemize}
\item Distributed Network: Blockchain is decentralized where data is not retained by any particular node, rather the data is made publicly available based on some agreed-upon decision process. As a result, data is always available and prevents one-point failure or data loss. Moreover, data alteration also gets difficult following this technology as all the participants acquire same copy of data.  
\item Shared Ledger: The participants in the network maintains a shared record of transactions known as a ledger which is public. Thus, blockchain becomes a trusted and transparent method of implementation. The participants run algorithms to measure the validity of an initiated transaction of a digital record and verify the planned dealing. If majority of the participants agree upon a common decision about the validity of the transaction, then the new transaction is included in the blockchain, recorded in the ledger and broadcast in the network for update.  
\item Digital Transactions: Transactions are structured into blocks. Each block contains a cryptologic hash to the previous block within the blockchain. This hash provides integrity of the digital record.
\item Consensus: It is a protocol where all the participants in the blockchain network agree about the validation of a transaction. Consensus prevents compromised participants problem as no single participant can take monopoly decision. 
\end{itemize}

\section{Proposed Approach}
\begin{figure*}
\centering
  \includegraphics[width=0.7\linewidth]{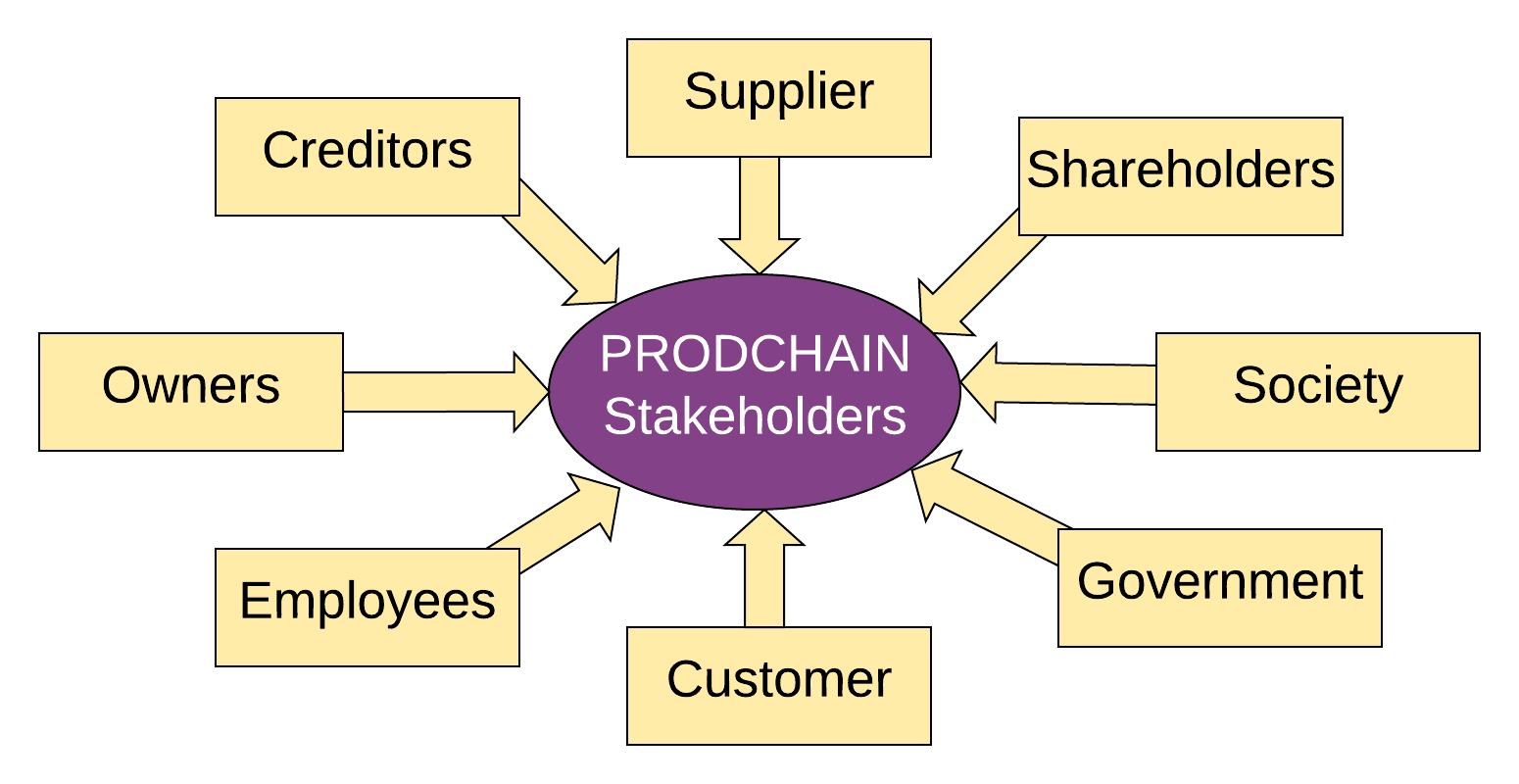}
  \caption{ PRODCHAIN stakeholders \cite{Mitchell}}
  \label{fig:fig02}
  \centering
\end{figure*}

The proposed solution PRODCHAIN aims to integrate the value chain (product chain) and supply chain with a blockchain backbone. Thus, an entirely transparent process will be available for the products and all the stakeholders are able to access the product information at any point of time being the part of the PRODCHAIN network. As this process will be beneficial for the stakeholders to manage their product from development to customer, it can also be considered as a manufacture-to-consumption chain. Figure \ref{fig:fig02} summarizes the basic stakeholders of common for product chain and supply chain that are considered for PRODCHAIN.

\begin{figure*}
\centering
  \includegraphics[width=0.8\linewidth]{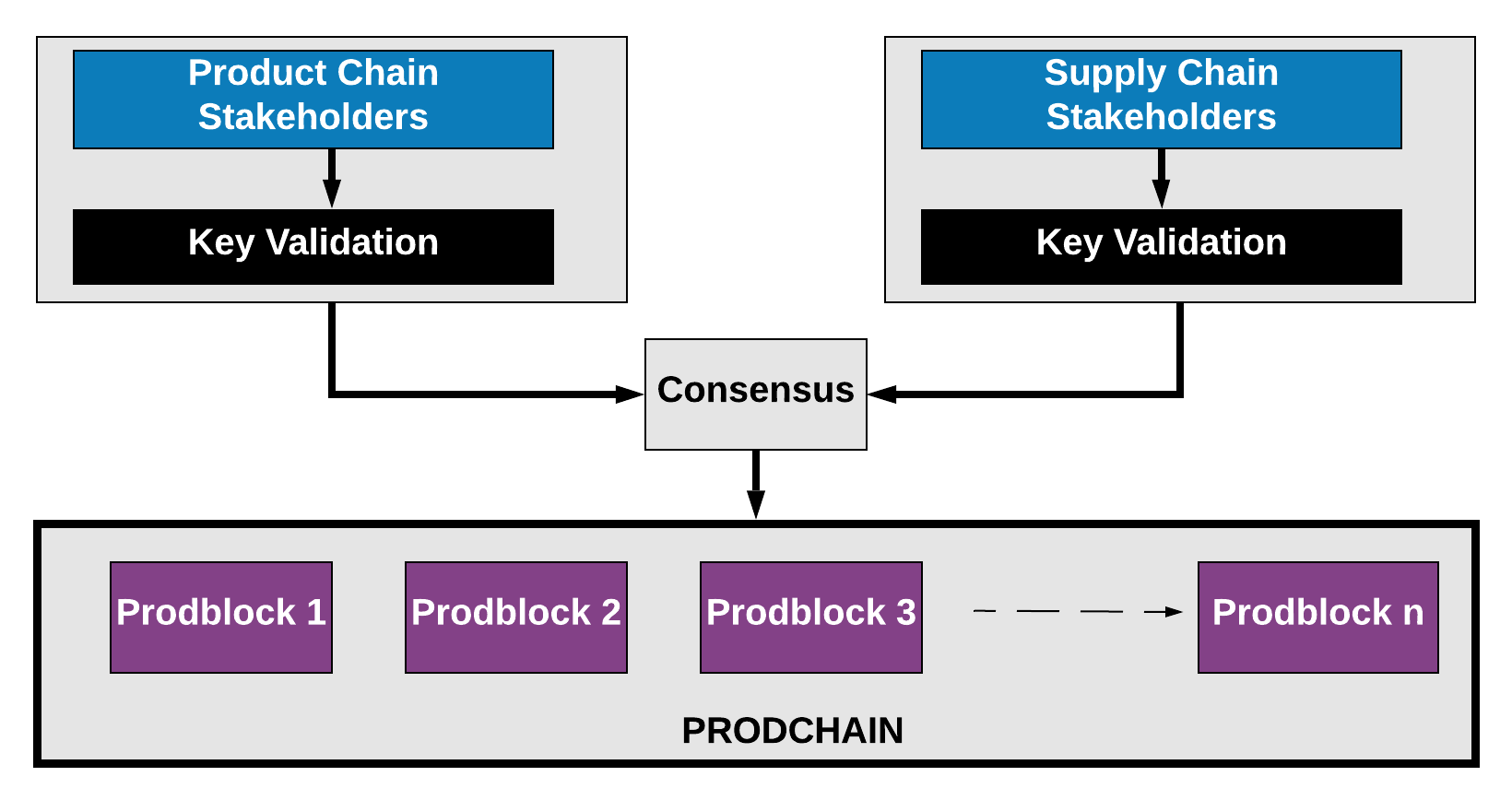}
  \caption{Proposed PRODCHAIN functioning}
  \label{fig:fig03}
  \centering
\end{figure*}

The PRODCHAIN nodes (users) register themselves with proper credentials. After successful validation of the credentials, the users are allowed to access the PRODCHAIN either to access the information or to publish the information.  The overall process is shown in Figure \ref{fig:fig03}. The following subsections explain the phases through which PRODCHAIN is established successfully.

\subsection{Identity management and key generation process}
\begin{figure*}
\centering
  \includegraphics[width=0.5\linewidth]{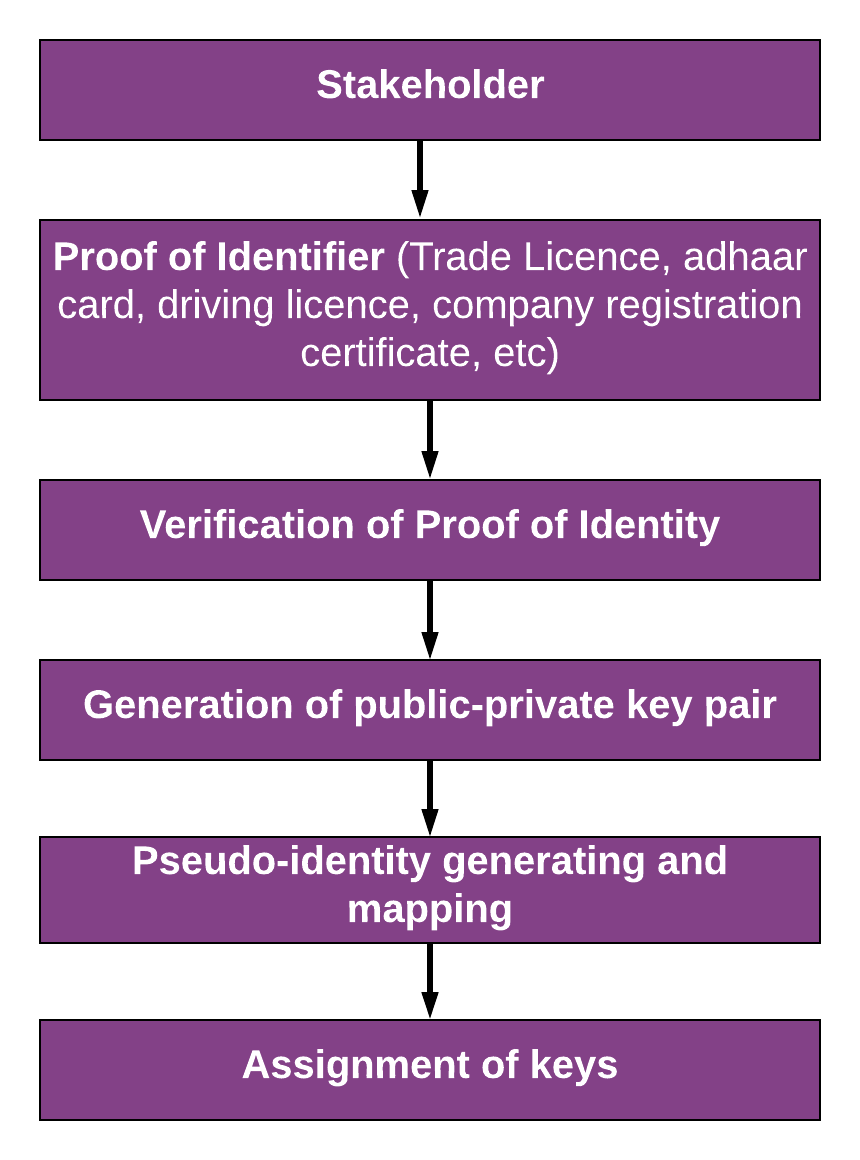}
  \caption{Stakeholder registration process}
  \label{fig:fig04}
  \centering
\end{figure*}

All the stakeholders are required to register for the PRODCHAIN network so that they can participate in initializing, validating and viewing the product block as per the requirement of the PRODCHAIN processes. The registration process also helps the stakeholders to activate their PRODCHAIN wallet that consists of a pseudo-identity, public-private key pair. These keys are further used for product block generation and identity verification process with signature. The registration process of the stakeholders is shown in Figure \ref{fig:fig04}. 

The registration process by the stakeholders starts with the submission of identity proof such as trade license, national/international identity documents. After the verification, a pseudo-identity is generated with that particular user id and public-private keys are generated for the user. For the generation of public-private key pairs and signcryption we have followed the bilinear pairings with some modification \cite{li2007efficient}. We have used the concept of symmetric random function as shown in \cite{saha2017symmetric} for generating this random basis \cite{neelamani2007nearly}. The random basis $\aleph$  is generated with Proof Metrics (PM) presented at the time of registration.  The partial private key lattice $\mathscr{K}$ is calculated using $\mathscr{K}_{part}$ by the key centre with the same PM. Both these lattices are used in the generation of public-private key pair by the key centre to use in lattice-based signcryption. The lattice generated by a set of vectors, $\mathscr{B}$, is given by:

\begin{equation}
Y=\Laplace{ (\mathscr{B}) }   =\{ \mathscr{B}_x = \sum_{i=1}^{m} \bar x \in \mathbb{Z}^m\}
\end{equation}

and where $\mathscr{B} = \{b_1,b_2,...b_m \}$. $\mathscr{B}$ is called the basis of lattice $Y$. The process of key pair generation is shown in Algorithm \ref{eqn01}:

\begin{algorithm}[h]
\caption{Key Generation\label{eqn01}}
\SetAlgoLined
\KwInput{ $(\aleph,\mathscr{K}) $    }
\KwOutput{Key pair $(K_{u+},K_{u-})$}
\v{b}: $G_1 \times G_2 \rightarrow \mathscr{R}_{q,[1]}$\\
$P:Gen (\mathscr{R}_{q,[1]} )$ such that $q \equiv 1 \pmod {2n}$\\
$K_{u-} \leftarrow \aleph(K_{part}) $\\
$K_{u+} \leftarrow P K_{u-} + \acute{b} $\\
\Return $K_{u+}$  and $K_{u-}$
\end{algorithm}

In this case, $(\aleph,\mathscr{K})$ is cyclic group $G_1,G_2$  of prime order $q \geq 2^k$. $K_{u+}$ is the public key and $K_{u+}$ is the private key. $R_{q,[1]} \in  G_1 \times G_2$ is the well-known ring $R_q= \mathbb{Z}_q  \frac{[X]}{(X_m+1)}$  with $q \equiv 1 \pmod{2n}$ and have the set of elements in the range of [-1,1] \cite{Lyubashevsky}. $X$ is a polynomial form of the lattice Y. Here, $\mathbb{Z}_q$ denotes residue class $\pmod q$. These keypairs $(K_{u+}$  and $K_{u-}$)  are used for signcryption/unsigncryption process by the users whenever they want to access the PRODCHAIN.  Furthermore, the pseudo-identity is provided by using Lattice Based Hash Function (LASH) with PM \cite{LASH}.

\subsection{Signcryption and Unsigncryption process}
Users get the keys from the key centre and performs the signcryption process for validation of the data block generated by the users. We have introduced a new term "PRODBLOCK" for this purpose. PRODBLOCK is a data block related with product information integrating the information from the value chain and supply chain. This product information includes data from product chain and also supply chain (For example: product manufacturer details, timestamps, quality status, warehouse status, shipment time, required customer information, and so on) and this is included in the $d_i$  part of $m_i:\{d_i,Iden_p,K_{u+}\}$. In this scenario, it is to be understood that signcryption process is executed by the PRODBLOCK initializer whereas the other users for that PRODBLOCK need to unsigncrypt the data before taking participation in consensus. Using the signcryption it provides confidentiality, integrity, authentication and non-repudiation services altogether and thus minimizing the computation cost as compared to the generic processes of the mentioned services one by one. The use of lattice-based cryptography for signcryption in blockchain is also advantageous for post-quantum cryptographic computation as lattice cryptographic processes are robust against quantum computed attacks so far. The signcryption and unsigncryption process is shown in Algorithm \ref{eqn02} and Algorithm \ref{eqn03} respectively.

The signcryption process uses three hashing processes in different steps. Hashing functions provide integrity of the data contained in a PRODBLOCK. Using the private key of the sender $K_{u-}$ for the hashed data provides signature. Therefore, the signcryption process is able to provide confidentiality, integrity, authentication, digital signature and non-repudiation simultaneously in a more effective way.

\begin{algorithm}[h]
\caption{Signcryption by PRODBLOCK initializer\label{eqn02}}
\SetAlgoLined
\KwInput{$(K_{u+},K_{u-}),m_i$}
\KwOutput{ciphertext c}
$r \in Z_q$\\
$R \in G_1$\\
$T=rP$\\
$H_1:G_1 \rightarrow \{0,1\}^n$ \\
$y= m_i  \oplus H_1 (R) $\\
$H_2: {\{0,1\}}^{n_1+(n+1)l} $ \\
$w= K_{u-} H_2(y,T,K_{r+1},K_{r+2},..K_{r+_N})$\\
$H_3: {G_1}^3 \rightarrow {\{0,1\}}^l$\\
$z_i= R \oplus H_3(T,K_{r+i},rTK_{r+i}) \forall i=1,2,..,N$\\
$ c=(T,y,w,z_i) $ \\
\Return c
\end{algorithm}

$K_{r+_i}$ represents the public key $K_{u+}$ of an individual user, $n$ is the number of bits, $l$ is an agreed upon public value Once the ciphertext c is computed, the PRODBLOCK initiator publishes the prodblock in the PRODCHAIN. Observing this attempt of publishing, the other users of the PRODCHAIN verify or unsigncrypt the ciphertext c.

\begin{algorithm}[h]
\caption{Unsigncryption process in PRODCHAIN\label{eqn03}}
\SetAlgoLined
\KwInput{$c,(K_{u+} ,K_{u-}) $}
\KwOutput{Decision on prodblock $m_i$}
$R= z_i \oplus H_3 (T,K_{r+i},K_{r-} T)$\\
$m_i= y \oplus H_3  (R)$\\
$ h= H_2(y,T,K_{r+1},K_{r+2},..K_{r+N} )$  \\
\eIf { $\acute{b}(P,w)= \acute{b}(K_{u+},h)$ }
{Publish $m_i$ in PRODCHAIN}
{Abort the transaction}

\end{algorithm}

$K_{r-i}$ is the private key of receiver $i$ and $K_{u+}$ represents the public key of the prodblock initiator. PRODCHAIN is an integrated proposed blockchain approach that combines the general value chain or product chain functionalities and supply chain activities on a public platform in distributed ledgers. But, before the prodblock publication in PRODCHAIN, the users (stakeholders) need to participate in a new consensus approach derived from smart contracts called of Proof-of-Accomplishment (PoA). This will help the PRODCHAIN to obtain an unbiased decision for a product information ($m_i$).

\subsection{Consensus: Proof-of-Accomplishment (PoA)}
We have derived the concept of PoA from the generic smart contracts \cite{Voshmgir}. Smart contracts are self-verifying, self-executing and tamper-resistant. Inheriting these features into PoA, these consensuses are able to provide increased degree of legal and contractual security and map the legal obligations automatically. Moreover, PoA is efficient in terms of transparency, less number of intermediaries and and reduces transactional cost. It is more beneficial in case of heterogeneous stakeholders. PoA is basically based on rating. We have used here 5-point rating scale to evaluate the anonymous decision through consensus. A product is basically started with a product chain and eventually proceeds for supply chain. Therefore, the environment of e-commerce involves heterogeneous stakeholders. To obtain a non-ambiguous and transparent decision based on consensus each stakeholder provides a rating to its service. For example, if a product has been ready and stored in the warehouse in a timely manner, the highest rating is to be used. Similarly, the delay will decrease the rating by 1. We have used delay factor $\Delta r=-1$, which means one day delay in production is going to decrease the rating by 1. Following the same process, other services such as shipment, transportation and delivery all the processes can have the corresponding rating. Once these ratings are visible in the PRODCHAIN, the intended stakeholder (user) is allowed to publish data information following some condition, else severity of monitoring can be imposed. The process of the PoA is summarized in Algorithm \ref{eqn04}.

\begin{algorithm}[h]
\caption{PoA contract\label{eqn04}}
\SetAlgoLined
\KwInput{Service time}
\KwOutput{PRODBLOCK access}
\par
Check service time (while rating i!=0)

\eIf {$service\_time > upper\_threshold$} {Decrement $r$ with $\Delta r$ where $r$ is the rating}
{
Set $r=5$\;
}
\par

Check the rating of the intended stakeholder\\
\eIf {$(r>0)$} 
{Allow access to PRODCHAIN publish}
{Severity-concern access}

\end{algorithm}

\section{Results and discussion}
The presented application framework of blockchain in value (product) chain-supply chain integration is new in the direction of distributed supply chain management. Therefore, the comparison of the experimented results has not been done in this paper. Table \ref{tab:tabA} describes the implementation framework. The implementation process, related results and theoretical advantages are explained in following subsections.


\begin{table*}

\caption{Implementation framework\label{tab:tabA}}{
\begin{tabular}{|l|l|}
\hline
Consensus protocol                   & Proof of Accomplishment (new introduction) \\
\hline
Geographic distribution of nodes     & Ethereum network, 20 nodes                 \\
\hline
Hardware environment of all peers    & 3.3 GHz, 8 GB RAM, Octa-core, 2 TB HDD     \\
\hline
Number of nodes involved in the test  transaction & 6                                          \\
\hline
Test tools and framework             & Hyperledger Caliper                        \\
\hline
Type of data store used              & CouchDB   \\ 
\hline
\end{tabular}
}
\end{table*}

\subsection{Implementation process}
\begin{figure*}
  \includegraphics[width=\linewidth]{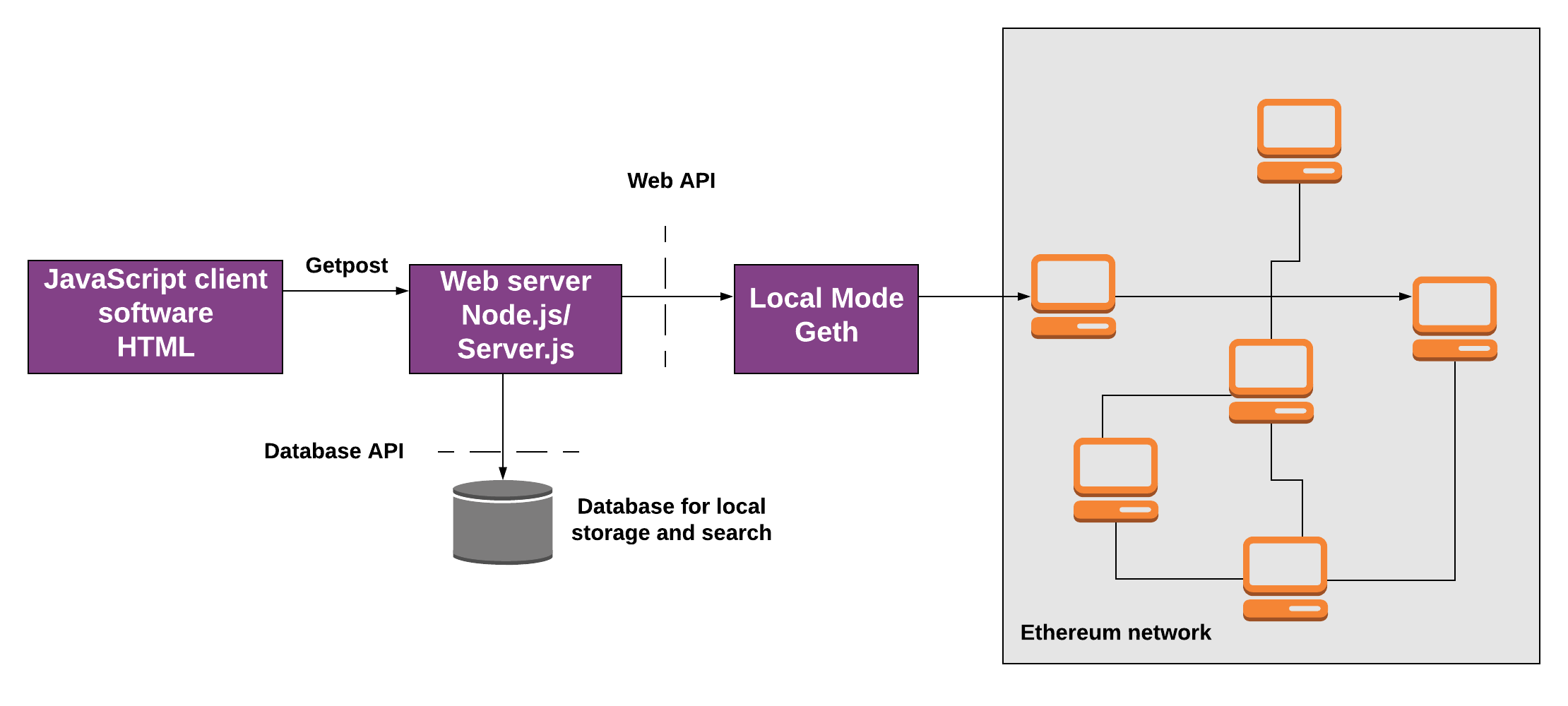}
  \caption{Implementation framework}
  \label{fig:fig05}
\end{figure*}

We have used Ethereum network with solidity contract and Remix IDE for the compilation of PoA. The architecture of the implementation is shown in Figure \ref{fig:fig05}. It shows that the implementation framework is consisted of clients having software for javascript and HTML. Webservers handle the client requests for blockchain and also uses a local database. Finally, Geth console is used to connect with the Ethereum network. The steps of implementation are:

\begin{enumerate}
\item Pre-installation of Homebrew and Node/npm.
\item Installation of Ethereum, Solidity, Remix IDE and Microsoft LatticeCrypto Library.
\item Genesis blocks are initialized.
\item PRODCHAIN is initialized with two blocks and three virtual organization accounts with wallets.
\item A folder is created for the blockchain to reside.
\item Private Ethereum Blockchain is initiated and run with lattice cryptographic signatures.
\item Geth Javascript console is used to connect to the private Ethereum blockchain.
\item Account has been created and dummy Ethers are mined.
\item PoA rating and decision condition are created in solidity and included in Ethereum. 
\item Remix IDE is initialized to deploy the generated PoA.
\item Remix IDE is updated with wallet account of the users and Ethereum network.
\item PoA is executed on Ethereum blockchain.
\end{enumerate}

\subsection{Results and discussion}
We have measured the system's performance to generate 100 prodblocks and have calculated the systems resource consumptions as shown in Table \ref{tab:tabB}. Various timing measurements are evaluated as per the definition in Table \ref{tab:tabB} and space complexity is also measured.

\begin{table*}

\caption{Timing measurements\label{tab:tabB}} {

\begin{tabular}{| p{3cm} | p{5cm} | p{5cm} |}
\hline
\textbf{Performance parameter}  & \textbf{Definition}   & \textbf{Formula for calculation used}      \\
\hline\hline
Read latency           & Time between a submission of a read request and receipt of a reply.  & Read Latency = Time when response received – submit time                              \\
\hline
Read Throughput        &  The number of read operations completed in a defined time period, expressed as reads per second (RPS).  & Read Throughput = Total read operations / total time in seconds                       \\
\hline
Transaction Latency    & Time taken for a transaction's effect to be usable across the network.  & Transaction Latency = (Confirmation time @ network threshold) – submit time           \\
\hline
Transaction Throughput & The rate at which valid transactions are committed by the blockchain in a defined time period. This rate is expressed as transactions per second (TPS) at a network size. & Throughput = Total committed transactions / total time in seconds @ \#committed nodes\\
\hline
\end{tabular}
}
\end{table*}

Transaction latency time starts from the time of submission to the point the result is widely available in the network. It includes the propagation time and any intermediating settling time due to the consensus mechanism in place. Transaction throughput is not considered for a single node rather, it is observed for the overall blockchain network for all the nodes to commit the transactions properly. 

Table \ref{tab:tab03} shows that the time consumption increases with the increasing number of block generation. The rate of increasing factor is linear and therefore it is efficient in the blockchain application paradigm. We have measured the commit time of all the 20 nodes as shown in Table \ref{tab:tab04}. The commit times help in calculating the Transactional latency and transactional throughput.

\begin{table*}[]
\label{Proposed Approach Performance}\label{tab:tab03}{

\begin{tabular}{|l|l|l|l|l|l|l|l|l|l|l|l|}
\hline
& 10                            & 20    & 30    & 40     & 50     & 60     & 70     & 80     & 90     & 100    & complexity             \\
\hline\hline
Read latency (seconds)        & 10.66 & 13.5  & 17.33  & 18.33  & 21.5   & 22.01  & 22.23  & 24.07  & 27.67  & 33.66      &  O(log n) \\
\hline
Transaction latency (seconds) & 46.43 & 78.70 & 111.33 & 173.33 & 208.23 & 260.47 & 317.63 & 352.03 & 371.33 & 410.02     & O(N log n)\\
\hline
\end{tabular}
}
\end{table*}

The  space complexity is $O(n)$. $n$ is the number of prodblocks, N is the number of users involved in a transaction of prodblock and m is the conditional decision in smart contract. The overall time is of generation of PRODBLOCK and publishing it in PRODCHAIN. Assumption of the internet speed is 120Kbps.

\begin{table*}[]
\caption{Commit time of node (s)\label{tab:tab04}}{
\begin{tabular}{|l|l|l|l|l|l|l|l|}
\hline
Node   & Commit time (s)  & Node    & Commit time (s)  & Node    & Commit time (s)  & Node    & Commit time (s)\\
\hline\hline
Node 1 & 6.43                  & Node 6  & 5.33                  & Node 11 & 6.67                  & Node 16 & 10.33                 \\
\hline
Node 2 & 7.33                  & Node 7  & 3.67                  & Node 12 & 7.01                  & Node 17 & 9.80                  \\
\hline
Node 3 & 4.66                  & Node 8  & 5.33                  & Node 13 & 6.50                  & Node 18 & 7.66                  \\
\hline
Node 4 & 4.00                  & Node 9  & 9.88                  & Node 14 & 7.00                  & Node 19 & 5.33                  \\
\hline
Node 5 & 6.00                  & Node 10 & 11.01                 & Node 15 & 10.33                 & Node 20 & 5.00  \\
\hline
\end{tabular}
}
\end{table*}

\begin{figure*}
\centering
  \includegraphics[width=0.6\linewidth]{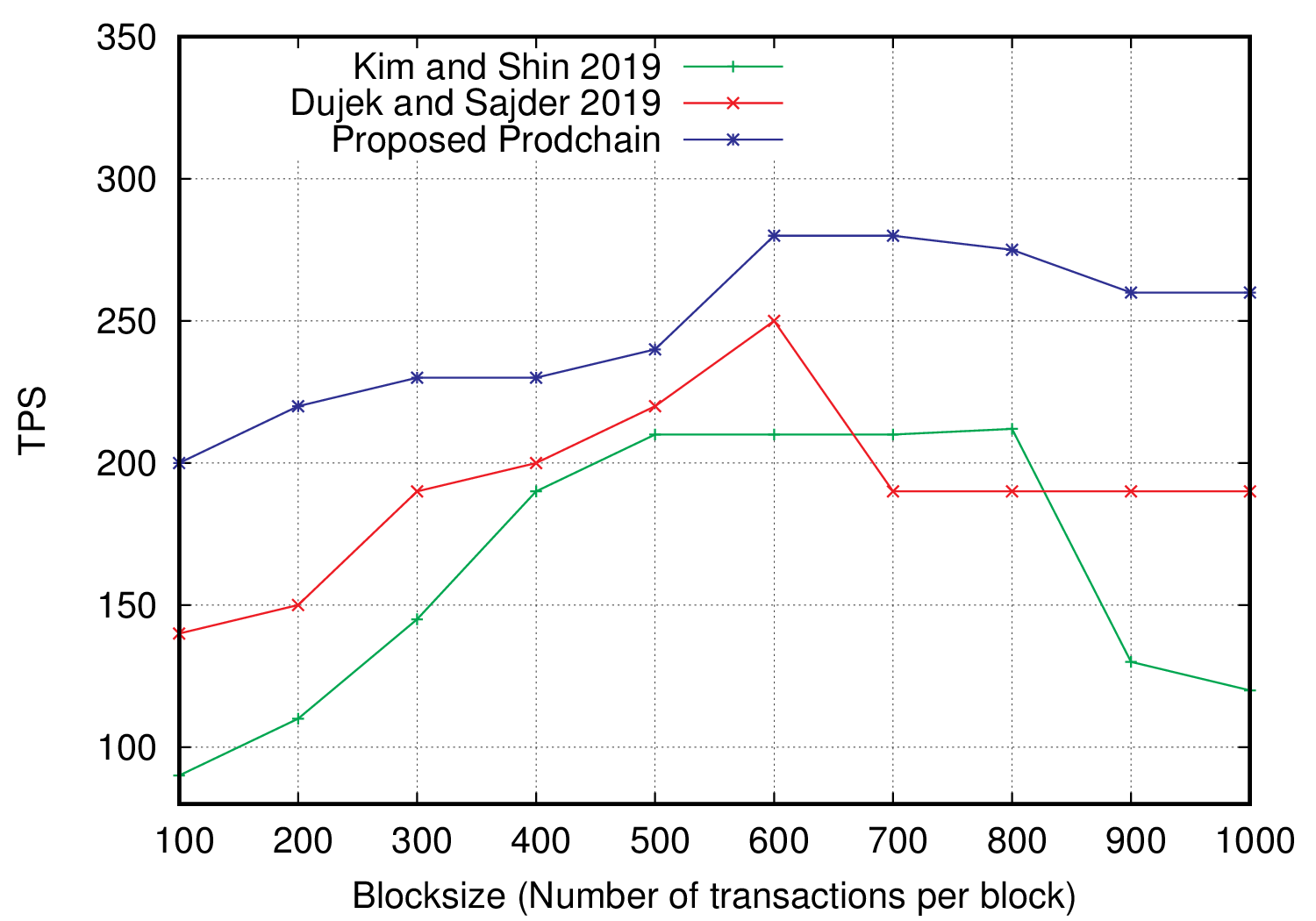}
  \caption{TPS vs Blocksize}
  \label{fig:fig06}
\centering
\end{figure*}

\begin{figure*}
\centering
  \includegraphics[width=0.6\linewidth]{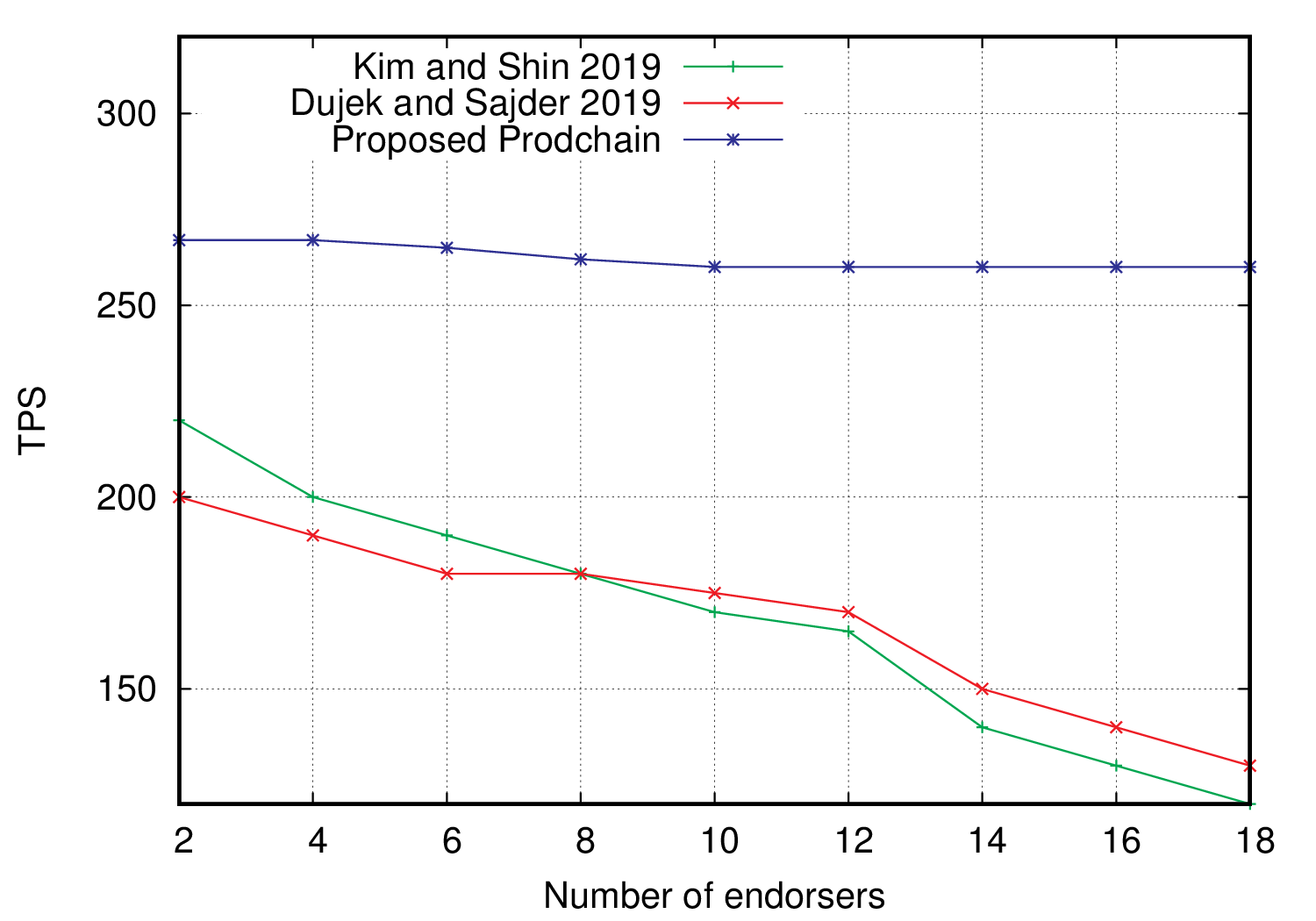}
  \caption{TPS vs Number of endorsers}
  \label{fig:fig07}
  \centering
\end{figure*}

We have also performed Hyperledger Caliper framework \cite{Caliper} to test the proposed blockchain solution. The results of transaction throughput measurements are given in Table \ref{tab:tab05}. It shows the measurements of Transaction throughput in the Caliper framework. It shows that the success rate of the prodblock generation is almost 100\% though with the increasing number of blocks success rates reduce by 0.33\% on average. The reduced success rates are occurred due to the problems such as consensus errors (e.g. endorsement policy not satisfied), syntax errors (e.g. invalid input, problem in signature, repeated transactions) and version errors. With the average block delay of approximately five prodblocks, Prodchain is efficient in processing blocks commitment. To evaluate the proposed system in depth, we have also measured the transaction throughput as a function of varying block size and number of endorsers in the network. The endorser simulates transactions and in turn prevents unstable or non-deterministic transactions from passing through the network. A transaction is sent to an endorser in the form of a transaction proposal. We have used variable blocksize by increasing the number of transactions per block from 1 to 1000 and number of endorsers are varied from 1 to 19 as we have used total 20 nodes in Ethereum network. We have compared the TPS results with other two recent blockchain applications as in \cite{Dujak2019} and \cite{KimShin2019}. The results are shown in Figure \ref{fig:fig06} and Figure \ref{fig:fig07}, respectively. 

Figure \ref{fig:fig06} shows that only 4.7\% of the transactions per block do have a negative impact on throughput with average decreasing of 2.3\%. Although, the throughput increases to 100 transactions per block, but degradation starts with the number of transactions increasing beyond 600. The maximum throughput of 289 tps lands around 250 tps with the proposed Prodchain. This parameter tells the blockchain orderer (Prodchain initiator) how many transactions that can be included inside a batch which is then sent to committers (receivers here) to form the next block. The other approaches in comparison have a degraded performance in issuing the blocks as tps reduces significantly which is approximately 30\% (\cite{KimShin2019}) and 28\% (\cite{Dujak2019}) less as compared to the proposed approach. Figure \ref{fig:fig07} shows that Prodchain is able to maintain the throughput stability of 260 tps with an increasing number of endorsers. Though the initialization with a smaller number of endorsers lead to the maximum throughput of 267 tps with the increasing number of endorsers affect the performance of prodchain very less with 2.62\% decreasing factor. The other two approaches shows a reduction of 33\% and 42\% in the same parameter. This makes Prodchain applicable and desirable for blockchain-based solutions as it is able to handle the scalability issues. The use of PoA and lattice-based cryptographic approaches make Prodchain efficient in processing and increased TPS as well. In reality, endorsers do not affect the Hyperledger performance, here we have used simultaneously for clients and endorsers and therefore sharing the resources which have shown an effect in degraded performance.

\begin{table*}
\caption{Results for Caliper framework\label{tab:tab05}}{
\begin{tabular}{|l|l|l|}
\hline
Number of prodblocks & Success rate & Transaction Throughput \\
\hline\hline
10                   & 100\%        & 7 blocks/sec           \\
\hline
20                   & 100\%        & 14 blocks/sec          \\
\hline
30                   & 100\%        & 24 blocks/sec          \\
\hline
40                   & 100\%        & 33 blocks/sec          \\
\hline
50                   & 100\%        & 41 blocks/sec          \\
\hline
60                   & 99.9\%       & 53 blocks/sec          \\
\hline
70                   & 99.87\%      & 62 blocks/sec          \\
\hline
80                   & 99.6\%       & 74 blocks/sec          \\
\hline
90                   & 99.5\%       & 81blocks/sec           \\
\hline
100                  & 99.48\%      & 95 blocks/sec \\   
\hline
\end{tabular}
}
\end{table*}


\subsection{Theoretical analysis}
We have analysed PRODCHAIN theoretically based on its functions and ability. We have identified the following major benefits.

\begin{itemize}
\item Re-engineering: Forward engineering is easy as compared to re-engineering or reverse engineering. PRODCHAIN provides a transparent platform of products about their development and supply. Therefore, re-engineering will be easier to identify a roll-back point in the total chain.  
\item Customer satisfaction: Customers generally concern for end products. Obtaining a clear view of product development, maintenance and supply customer will be more satisfied with the transparency of the system. 
\item Trust: PRODCHAIN improves the trust factor among the stakeholders. Basically, non-repudiation is well achieved as it applies consensus so that there will be no single point of authority in the PRODCHAIN network to take the decision.  
\item Measurability: The proposed solution also helps in measuring a product in a multidimensional way using location-based analysis, feature-based analysis and prediction analysis. Moreover, it is a transparent way to monitor whether an intermediary is holding the supply to resupply with high cost, thus avoiding surplus or dearth for any product.  
\item Bridged platform: PRODCHAIN creates a logical bridge between the product chain and supply chain. This bridge helps to avoid the anomaly intermediaries so that manufacturing-to-consumption can be executed smoothly. The distributed nature, openness and immutable features help PRODCHAIN to prevent any data manipulation. 
\item Security services: The use of keys and block hashes (like in general blockchain approach) in PRODCHAIN helps to provide confidentiality, authentication and non-repudiation through digital signatures, the integrity of prodblock by using crypto hash and thus PRODCHAIN is secure.  
\item Data availability: PRODCHAIN is distributed. Therefore, even in the presence of maliciousness data will be available in the PRODCHAIN network for the legitimate access of the stakeholders. 
\item Stakeholders perspective: All the stakeholders are able to get a view of their product with all corresponding information. The information is accurate as the other stakeholders have verified the information with a consensus protocol and thus biasness or data infiltration is prevented. 
\end{itemize}

All the above benefits are significant for PRODCHAIN and thus making it logically and practically efficient to be implemented with cloud computing.

\section{Conclusion}
A blockchain-based solution (PRODCHAIN) that integrates product/value and supply chains has been developed for squashing the manipulation of data, impact of monopoly, unbalanced product price, uncertainty of product quality, duplicate products in the \textit{e}-commerce world. The method is logically derived for implementation in Ethereum Base. The solution provides a transparent view of the data, starting from the development to the consumption of the products.  Theoretically,  the method is able to change the present product development and consumption scenario efficiently and securely from all perspectives. The experimental results also support the theoretical finding and confirm the efficiency of PRODCHAIN in the field of e-commerce.  In short, this generic blockchain framework is beneficial for any product management systems.     

\bibliography{main}

\begin{thebibliography}{10}
\providecommand{\url}[1]{#1}
\csname url@samestyle\endcsname
\providecommand{\newblock}{\relax}
\providecommand{\bibinfo}[2]{#2}
\providecommand{\BIBentrySTDinterwordspacing}{\spaceskip=0pt\relax}
\providecommand{\BIBentryALTinterwordstretchfactor}{4}
\providecommand{\BIBentryALTinterwordspacing}{\spaceskip=\fontdimen2\font plus
\BIBentryALTinterwordstretchfactor\fontdimen3\font minus
  \fontdimen4\font\relax}
\providecommand{\BIBforeignlanguage}[2]{{%
\expandafter\ifx\csname l@#1\endcsname\relax
\typeout{** WARNING: IEEEtran.bst: No hyphenation pattern has been}%
\typeout{** loaded for the language `#1'. Using the pattern for}%
\typeout{** the default language instead.}%
\else
\language=\csname l@#1\endcsname
\fi
#2}}
\providecommand{\BIBdecl}{\relax}
\BIBdecl

\bibitem{wong2019critical}
D.~T. Wong and E.~W. Ngai, ``Critical review of supply chain innovation
  research (1999--2016),'' \emph{Industrial Marketing Management}, 2019.

\bibitem{REDDYK201940}
\BIBentryALTinterwordspacing
J.~M.~R. K, N.~R. A, and K.~L, ``A review on supply chain performance
  measurement systems,'' \emph{Procedia Manufacturing}, vol.~30, pp. 40 -- 47,
  2019, digital Manufacturing Transforming Industry Towards Sustainable Growth.
  [Online]. Available:
  \url{http://www.sciencedirect.com/science/article/pii/S235197891930037X}
\BIBentrySTDinterwordspacing

\bibitem{moretto2019supply}
A.~Moretto, L.~Grassi, F.~Caniato, M.~Giorgino, and S.~Ronchi, ``Supply chain
  finance: From traditional to supply chain credit rating,'' \emph{Journal of
  Purchasing and Supply Management}, vol.~25, no.~2, pp. 197--217, 2019.

\bibitem{xu2019supply}
M.~Xu, Y.~Cui, M.~Hu, X.~Xu, Z.~Zhang, S.~Liang, and S.~Qu, ``Supply chain
  sustainability risk and assessment,'' \emph{Journal of Cleaner Production},
  vol. 225, pp. 857--867, 2019.

\bibitem{braz2018bullwhip}
A.~C. Braz, A.~M. De~Mello, L.~A. de~Vasconcelos~Gomes, and P.~T.
  de~Souza~Nascimento, ``The bullwhip effect in closed-loop supply chains: A
  systematic literature review,'' \emph{Journal of cleaner production}, vol.
  202, pp. 376--389, 2018.

\bibitem{schniederjans2019supply}
D.~G. Schniederjans, C.~Curado, and M.~Khalajhedayati, ``Supply chain
  digitisation trends: An integration of knowledge management,''
  \emph{International Journal of Production Economics}, 2019.

\bibitem{koberg2018systematic}
E.~Koberg and A.~Longoni, ``A systematic review of sustainable supply chain
  management in global supply chains,'' \emph{Journal of cleaner production},
  2018.

\bibitem{Wailgum}
T.~Wailgum, ``{Fraud, theft risks in supply chains are everywhere},''
  \url{https://www.networkworld.com/article/2278876/fraud--theft-risks-in-supply-chains-are-everywhere.html},
  2008.

\bibitem{Pomerantz}
P.~L.~W. Glenn~Pomerantz, Jeff~Cascini, ``{Supply Chain Fraud: Risk Management
  for Retail \& Consumer Products Companies},''
  \url{https://www.bdo.com/blogs/consumer-business-compass/december-2018/supply-chain-fraud},
  2018.

\bibitem{behzadi2018agribusiness}
G.~Behzadi, M.~J. O’Sullivan, T.~L. Olsen, and A.~Zhang, ``Agribusiness
  supply chain risk management: A review of quantitative decision models,''
  \emph{Omega}, vol.~79, pp. 21--42, 2018.

\bibitem{jonkman2019integrating}
J.~Jonkman, A.~P. Barbosa-P{\'o}voa, and J.~M. Bloemhof, ``Integrating
  harvesting decisions in the design of agro-food supply chains,''
  \emph{European Journal of Operational Research}, vol. 276, no.~1, pp.
  247--258, 2019.

\bibitem{duijzer2018literature}
L.~E. Duijzer, W.~van Jaarsveld, and R.~Dekker, ``Literature review: The
  vaccine supply chain,'' \emph{European Journal of Operational Research}, vol.
  268, no.~1, pp. 174--192, 2018.

\bibitem{wen2018fashion}
X.~Wen, T.-M. Choi, and S.-H. Chung, ``Fashion retail supply chain management:
  A review of operational models,'' \emph{International Journal of Production
  Economics}, 2018.

\bibitem{megahed2018tactical}
A.~Megahed and M.~Goetschalckx, ``Tactical supply chain planning under
  uncertainty with an application in the wind turbines industry,''
  \emph{Computers \& Operations Research}, vol. 100, pp. 287--300, 2018.

\bibitem{Khos2019}
\BIBentryALTinterwordspacing
F.~Khosravi and G.~Izbirak, ``A stakeholder perspective of social
  sustainability measurement in healthcare supply chain management,''
  \emph{Sustainable Cities and Society}, vol.~50, p. 101681, 2019. [Online].
  Available:
  \url{http://www.sciencedirect.com/science/article/pii/S2210670719309576}
\BIBentrySTDinterwordspacing

\bibitem{cole2019role}
R.~Cole and J.~Aitken, ``The role of intermediaries in establishing a
  sustainable supply chain,'' \emph{Journal of Purchasing and Supply
  Management}, p. 100533, 2019.

\bibitem{Drahokoupil}
J.~Drahokoupil, ``{Production Chain, Economics},'' \url{
  https://www.britannica.com/topic/production-chain}, 2015.

\bibitem{Tarver}
E.~Tarver, ``{Value Chain vs. Supply Chain: What\'s the Difference?}'' \url{
  https://www.investopedia.com/ask/answers/043015/what-difference-between-value-chain-and-supply-chain.asp},
  2019.

\bibitem{daud2011lean}
A.~Daud and S.~Zailani, ``Lean supply chain practices and performance in the
  context of malaysia,'' \emph{Supply Chain Management--Pathways for Research
  and Practice}, p.~1, 2011.

\bibitem{World}
W.~Bank, ``{World Trade Organization. Global Value Chain Development Report
  2019: Technological Innovation, Supply Chain Trade, and Workers in a
  Globalized World (English). Washington, D.C.: World Bank Group},''
  \url{http://documents.worldbank.org/curated/en/384161555079173489/
  Global-Value-Chain-Development-Report-2019-Technological-\\Innovation-Supply-Chain-
  Trade-and- Workers-in-a- Globalized-World}, 2019.

\bibitem{montecchi2019s}
M.~Montecchi, K.~Plangger, and M.~Etter, ``It’s real, trust me! establishing
  supply chain provenance using blockchain,'' \emph{Business Horizons},
  vol.~62, no.~3, pp. 283--293, 2019.

\bibitem{Alam2020}
F.~A. Khan, M.~Asif, A.~Ahmad, M.~Alharbi, and H.~Aljuaid, ``Blockchain
  technology, improvement suggestions, security challenges on smart grid and
  its application in healthcare for sustainable development,''
  \emph{Sustainable Cities and Society}, vol.~55, p. 102018, 2020.

\bibitem{chang2019supply}
S.~E. Chang, Y.-C. Chen, and M.-F. Lu, ``Supply chain re-engineering using
  blockchain technology: A case of smart contract based tracking process,''
  \emph{Technological Forecasting and Social Change}, vol. 144, pp. 1--11,
  2019.

\bibitem{banerjee2018blockchain}
A.~Banerjee, ``Blockchain technology: supply chain insights from erp,'' in
  \emph{Advances in Computers}.\hskip 1em plus 0.5em minus 0.4em\relax
  Elsevier, 2018, vol. 111, pp. 69--98.

\bibitem{choi2019mean}
T.-M. Choi, X.~Wen, X.~Sun, and S.-H. Chung, ``The mean-variance approach for
  global supply chain risk analysis with air logistics in the blockchain
  technology era,'' \emph{Transportation Research Part E: Logistics and
  Transportation Review}, vol. 127, pp. 178--191, 2019.

\bibitem{choi2019blockchain}
T.-M. Choi, ``Blockchain-technology-supported platforms for diamond
  authentication and certification in luxury supply chains,''
  \emph{Transportation Research Part E: Logistics and Transportation Review},
  vol. 128, pp. 17--29, 2019.

\bibitem{behnke2019boundary}
K.~Behnke and M.~Janssen, ``Boundary conditions for traceability in food supply
  chains using blockchain technology,'' \emph{International Journal of
  Information Management}, 2019.

\bibitem{kamble2019modeling}
S.~S. Kamble, A.~Gunasekaran, and R.~Sharma, ``Modeling the blockchain enabled
  traceability in agriculture supply chain,'' \emph{International Journal of
  Information Management}, 2019.

\bibitem{AZZI2019582}
\BIBentryALTinterwordspacing
R.~Azzi, R.~K. Chamoun, and M.~Sokhn, ``The power of a blockchain-based supply
  chain,'' \emph{Computers \& Industrial Engineering}, vol. 135, pp. 582 --
  592, 2019. [Online]. Available:
  \url{http://www.sciencedirect.com/science/article/pii/S0360835219303729}
\BIBentrySTDinterwordspacing

\bibitem{wang2019making}
Y.~Wang, M.~Singgih, J.~Wang, and M.~Rit, ``Making sense of blockchain
  technology: How will it transform supply chains?'' \emph{International
  Journal of Production Economics}, vol. 211, pp. 221--236, 2019.

\bibitem{zhao2019blockchain}
G.~Zhao, S.~Liu, C.~Lopez, H.~Lu, S.~Elgueta, H.~Chen, and B.~M. Boshkoska,
  ``Blockchain technology in agri-food value chain management: A synthesis of
  applications, challenges and future research directions,'' \emph{Computers in
  Industry}, vol. 109, pp. 83--99, 2019.

\bibitem{huang2019use}
J.~Huang, S.~Li, and M.~Th{\"u}rer, ``On the use of blockchain in industrial
  product service systems: A critical review and analysis,'' \emph{Procedia
  CIRP}, vol.~83, pp. 552--556, 2019.

\bibitem{li2019blockchain}
M.~Li, L.~Shen, and G.~Q. Huang, ``Blockchain-enabled workflow operating system
  for logistics resources sharing in e-commerce logistics real estate
  service,'' \emph{Computers \& Industrial Engineering}, vol. 135, pp.
  950--969, 2019.

\bibitem{Lu2017}
Q.~{Lu} and X.~{Xu}, ``Adaptable blockchain-based systems: A case study for
  product traceability,'' \emph{IEEE Software}, vol.~34, no.~6, pp. 21--27,
  November 2017.

\bibitem{Jabbari}
P.~K. Arman~Jabbari, ``{Blockchain and Supply Chain Management},''
  \emph{Technical Report, University of California, Berkeley}, pp. 1--13, 2018.

\bibitem{Dujak2019}
S.~D. Dujak~D., ``Blockchain applications in supply chain,'' \emph{Kawa A.,
  Maryniak A. (eds) SMART Supply Network. EcoProduction (Environmental Issues
  in Logistics and Manufacturing}, pp. 21--46, 2019.

\bibitem{Ven2020}
V.~Venkatesh, K.~Kang, B.~Wang, R.~Y. Zhong, and A.~Zhang, ``System
  architecture for blockchain based transparency of supply chain social
  sustainability,'' \emph{Robotics and Computer-Integrated Manufacturing},
  vol.~63, p. 101896, 2020.

\bibitem{Longo2019}
F.~Longo, L.~Nicoletti, A.~Padovano, G.~d'Atri, and M.~Forte,
  ``Blockchain-enabled supply chain: An experimental study,'' \emph{Computers
  \& Industrial Engineering}, vol. 136, pp. 57 -- 69, 2019.

\bibitem{Helo2019}
P.~Helo and Y.~Hao, ``Blockchains in operations and supply chains: A model and
  reference implementation,'' \emph{Computers \& Industrial Engineering}, vol.
  136, pp. 242 -- 251, 2019.

\bibitem{Kou2018}
J.~S. Mahtab~Kouhizadeh, ``Blockchain practices, potentials, and perspectives
  in greening supply chains,'' \emph{Sustainability}, vol.~10, no.~10, pp.
  1--16, 2018.

\bibitem{Kshetri2018}
E.~L. N.~Kshetri, ``Blockchain adoption in supply chain networks in asia,''
  \emph{IT Professional}, vol.~21, no.~1, pp. 11--15, 2019.

\bibitem{Ying2018}
W.~Ying, S.~Jia, and W.~Du, ``Digital enablement of blockchain: Evidence from
  hna group,'' \emph{International Journal of Information Management}, vol.~39,
  pp. 1--4, 2018.

\bibitem{Tian2016}
F.~Tian, ``An agri-food supply chain traceability system for china based on
  rfid blockchain technology,'' in \emph{2016 13th International Conference on
  Service Systems and Service Management (ICSSSM)}, 2016, pp. 1--6.

\bibitem{Niko2018}
L.~J. William~Nikolakis and H.~Krishnan, ``How blockchain can shape sustainable
  global value chains: An evidence, verifiability, and enforceability (eve)
  framework,'' \emph{Sustainability}, vol.~10, pp. 1--16, 2018.

\bibitem{KimShin2019}
N.~S. Joon-Seok~Kim, ``The impact of blockchain technology application on
  supply chain partnership and performance,'' \emph{Sustainability}, vol.~11,
  no.~21, pp. 1--17, 2019.

\bibitem{Las2018}
M.~L. Henry M.~Kim, ``Toward an ontology‐driven blockchain design for
  supply‐chain provenance,'' \emph{Intelligent Systems in Accounting, Finance
  and Management}, vol.~25, no.~1, pp. 18--27, 2018.

\bibitem{Wang2020}
Z.~Wang, T.~Wang, H.~Hu, J.~Gong, X.~Ren, and Q.~Xiao, ``Blockchain-based
  framework for improving supply chain traceability and information sharing in
  precast construction,'' \emph{Automation in Construction}, vol. 111, p.
  103063, 2020.

\bibitem{Helo2020}
P.~Helo and A.~Shamsuzzoha, ``Real-time supply chain—a blockchain
  architecture for project deliveries,'' \emph{Robotics and Computer-Integrated
  Manufacturing}, vol.~63, p. 101909, 2020.

\bibitem{Liu2020}
Z.~Liu and Z.~Li, ``A blockchain-based framework of cross-border e-commerce
  supply chain,'' \emph{International Journal of Information Management},
  vol.~52, p. 102059, 2020.

\bibitem{Chen2020}
T.~Chen and D.~Wang, ``Combined application of blockchain technology in
  fractional calculus model of supply chain financial system,'' \emph{Chaos,
  Solitons \& Fractals}, vol. 131, p. 109461, 2020.

\bibitem{leewayhertz}
leewayhertz.com, ``{Blockchain Development Key Concepts},''
  \url{https://www.leewayhertz.com/blockchain-development-key-concepts/}, 2019.

\bibitem{Mitchell}
R.~Mitchell, ``{Salience Modelsupply Chain Management},''
  \url{https://www.slideshare.net/archanabinoy143/salience-modelsupply-chain-management},
  2019.

\bibitem{li2007efficient}
F.~Li, Y.~Hu, and S.~Liu, ``Efficient and provably secure multi-recipient
  signcryption from bilinear pairings,'' \emph{Wuhan University Journal of
  Natural Sciences}, vol.~12, no.~1, pp. 17--20, 2007.

\bibitem{saha2017symmetric}
R.~Saha and G.~Geetha, ``Symmetric random function generator (srfg): A novel
  cryptographic primitive for designing fast and robust algorithms,''
  \emph{Chaos, Solitons \& Fractals}, vol. 104, pp. 371--377, 2017.

\bibitem{neelamani2007nearly}
R.~Neelamani, S.~Dash, and R.~G. Baraniuk, ``On nearly orthogonal lattice bases
  and random lattices,'' \emph{SIAM Journal on Discrete Mathematics}, vol.~21,
  no.~1, pp. 199--219, 2007.

\bibitem{Lyubashevsky}
\BIBentryALTinterwordspacing
V.~Lyubashevsky, C.~Peikert, and O.~Regev, ``On ideal lattices and learning
  with errors over rings,'' \emph{J. ACM}, vol.~60, no.~6, pp. 43:1--43:35,
  Nov. 2013. [Online]. Available: \url{http://doi.acm.org/10.1145/2535925}
\BIBentrySTDinterwordspacing

\bibitem{LASH}
R.~Steinfeld, S.~Contini, K.~Matusiewicz, J.~Pieprzyk, J.~Guo, S.~Ling, and
  H.~Wang, ``Cryptanalysis of lash,'' in \emph{Fast Software Encryption},
  K.~Nyberg, Ed.\hskip 1em plus 0.5em minus 0.4em\relax Berlin, Heidelberg:
  Springer Berlin Heidelberg, 2008, pp. 207--223.

\bibitem{Voshmgir}
S.~Voshmgir, ``{Smart contracts, Token Economy},''
  \url{https://blockchainhub.net/smart-contracts/}, 2019.

\bibitem{Caliper}
Hyperledger, ``{Hyperledger Caliper},''
  \url{https://hyperledger.github.io/caliper/}, 2019.

\end{thebibliography}

\bibliographystyle{IEEEtran}

\end{document}